\documentclass[aps,pre,twocolumn,groupedaddress]{revtex4-1}

\usepackage{amsmath}
\usepackage{graphicx}
\usepackage{float}
\usepackage{siunitx}
\usepackage{xcolor}
\definecolor{darkgreen}{rgb}{0,0.6,0.0}

\begin{document}

\title{Run-and-tumble bacteria slowly approaching the diffusive regime}

\author{Andrea Villa-Torrealba}
\email[]{aavillat@gmail.com}
\author{Crist\'obal Ch\'avez Raby}
\author{Pablo de Castro}
\email[]{pdecastro@ing.uchile.cl}
\author{Rodrigo Soto}
\email[]{rsoto@dfi.uchile.cl}

\affiliation{Departamento de F\'isica, Facultad de Ciencias F\'isicas y Matem\'aticas, Universidad de Chile, Avenida Blanco Encalada 2008, Santiago, Chile}

\date{\today}

\begin{abstract}
The run-and-tumble (RT) dynamics followed by bacterial swimmers gives rise first to a ballistic motion due to their persistence, and later, through consecutive tumbles, to a diffusive process. 
	Here we investigate how long it takes for a dilute swimmer suspension to reach the diffusive regime as well as what is the amplitude of the deviations from the diffusive dynamics. A linear time dependence of the mean-squared displacement (MSD) is insufficient to characterize diffusion and thus we also focus on the excess kurtosis of the displacement distribution. Four swimming strategies are considered: (i) the conventional RT model with complete reorientation after tumbling, (ii) the case of partial reorientation, characterized by a distribution of tumbling angles, (iii) a run-and-reverse model with rotational diffusion, and (iv) a RT particle where the tumbling rate depends on the stochastic concentration of an internal protein.
	By analyzing the associated kinetic equations for the probability density function and simulating the models, we find that for models (ii), (iii), and (iv) the relaxation to diffusion can take much longer than the mean time between tumble events, evidencing the existence of large tails in the particle displacements. Moreover, the excess kurtosis can assume large positive values. In model (ii) it is possible for some distributions of tumbling angles that the MSD reaches a linear time dependence but, still, the dynamics remains non-Gaussian for long times. This is also the case in model (iii) for small rotational diffusivity. For all models, the long-time diffusion coefficients are also obtained.
	The theoretical approach, which relies on eigenvalue and angular Fourier expansions of the van Hove function, is in excellent agreement with the simulations.
\end{abstract}

\maketitle

\section{Introduction}
\label{Intro}
There are billions of different species of bacteria on Earth \cite{larsen2017inordinate}. Because of adaption, their life and swimming styles vary across a multitude of distinct environments and conditions \cite{whitman1998prokaryotes,louca2019census,pohl2017inferring,detcheverry2017generalized,seyrich2018statistical}. The vast majority have never been researched, and are thus dubbed \textit{Microbial Dark Matter}~\cite{hatfull2015dark}. On the other hand, the \textit{E.~coli} bacteria continue to be extensively studied. Their motion is usually modeled as a run-and-tumble (RT) dynamics. In fact, their flagella can rotate and propel the cell body in a ``run'' mode which can suddenly terminate whenever some of them reverse direction \cite{taktikos2013motility}. This leads to a quick reorientation mode called ``tumble''---which is then followed by another run---with an average tumbling angle of approximately $70^{\circ}$ \cite{berg1993random}. In the case of marine bacteria, up to 70\% of them are thought to have a distribution of tumbling angles peaked around $180^{\circ}$ instead \cite{johansen2002variability}. Examples include \textit{S.\ putrefaciens} and \textit{P.\ haloplanktis} \cite{barbara2003bacterial}, and thus in this case we can speak of a run-and-reverse motion.

In his seminal work \cite{berg1972chemotaxis,berg2008coli}, Berg showed that bacteria and other microswimmers performing run-and-tumble motion develop, in the long term, a diffusive motion. If $V$ is the characteristic run velocity and $\nu_0$ the tumble rate (or rotational diffusion coefficient, in the case of mutant swimmers that tumble only very rarely), the diffusion coefficient scales as $D\sim V^2/\nu_0$, with a prefactor that depends on the tumble properties. For example, in the case of three-dimensional Markovian swimmers, i.e., each tumble is uncorrelated from previous ones and tumble events are distributed as a Poisson process, $D=V^2/[3\nu_0(1-\langle\cos\theta_{\mathrm{s}}\rangle)]$, where $\theta_{\mathrm{s}}$ is the tumbling (or ``scattering'') angle between the pre- and post-tumble directors~\cite{berg1972chemotaxis,berg2008coli}. In the case of \textit{E. coli}, the data in Ref.~\cite{berg1972chemotaxis} gives
 $\langle\cos\theta_{\mathrm{s}}\rangle \simeq 0.33$,  $\nu_0\simeq\SI{1.2}{\second^{-1}}$, $V\simeq \SI{14.2}{\micro\meter/\second}$, which results in $D \simeq \SI{87}{\micro\meter^2/\second}$~\cite{lovely1975statistical}.
The diffusive description of bacterial spreading is extensively used because of its simplicity, which allows, for example, to couple this random dynamics with hydrodynamic flows and with the diffusion of nutrients and other chemicals, or to consider complex geometrical restrictions (for recent applications, see~\cite{angelani2014first,caprini2019transport,wagner2017steady,seyrich2019traveling,wagner2017steady}). Also, it is possible to include cell division and death by employing reaction--diffusion equations, as it is common in chemical and environmental engineering to describe the spatiotemporal spreading of bacteria~\cite{Gourley2004,britton1986,Messoud2013}. 
Finally, nonlinear effects as a density-dependent diffusion coefficient are key to describe motility-induced phase separation~\cite{tailleur2008statistical,cates2015motility}.
At short times, on the other hand, the swimmers' persistent motion gives rise to a ballistic motion. Na\"ively, the crossover time $T_\text{cross}$ between the ballistic and diffusive regimes is expected to be relatively small and to scale as $\nu_0^{-1}$. In this article we thoroughly show that, depending on tumbling strategies and parameters, the prefactor of this scaling can be quite large and thus the non-diffusive regime can persist for long times. This can happen even if the mean-squared displacement ($\rm{MSD}$) reaches a linear time dependence relatively quickly since having $\rm{MSD}\sim t$ is a necessary but not sufficient condition for being in the diffusive regime. Note that, associated with $T_\text{cross}$, there is a spatial scale $L_\text{cross}=\sqrt{{\rm MSD}(T_\text{cross})}$ where the diffusive description is not valid. Simulations of active Brownian particles (ABPs) with large values of $L_\text{cross}$ show that for lengths of this order a non-diffusive regime indeed arises \cite{caprini2019transport}.

Our motivation is to quantitatively study the dispersal process of bacteria. With that purpose in mind, we consider several run-and-tumble models which are distinct in swimming strategy and compare how slowly these microswimmers approach the diffusive regime. We also provide the spreading dynamics for temporal and spatial scales smaller than $T_\text{cross}$ and $L_\text{cross}$, respectively.
The swimming strategies considered here are different not only in terms of the distribution of tumbling angles but also in whether or not the tumbling rate remains constant over time. In particular we consider the Tu--Grinstein model \cite{tu2005white}, where the concentration of a phosphorylated internal protein named CheY-P changes stochastically with time \cite{dev2019run}, affecting the tumbling rate exponentially. 
Previous studies have discussed departures from diffusion by using the MSD for run-and-tumble swimmers \cite{shaebani2019transient} and through the excess kurtosis of the displacement distribution for ABPs \cite{ten2011brownian,zheng2013non,basu2018active}. More recently,
Ref.~\cite{put2019non} has studied the non-Gaussian behavior of interacting run-and-tumble particles in the context of active polymer chains and lattice models, where the authors considered simpler tumbling processes and employed analytical methods which are based on solving the associated Langevin equation. In the case of the present work, our analysis is done by performing simulations and derivations of both the MSD and the excess kurtosis for the different RT models, aiming to appropriately determine how long the system takes to reach the diffusive regime. Furthermore, the analytical part is carried out from associated kinetic equations, with Fokker-Planck terms to describe rotational diffusion and the evolution of the protein concentration \cite{martens2012probability} coupled with a Lorentz term to account for the tumbling \cite{saintillan2010dilute,Saintillan2018,soto2016kinetic}. The simulations are essentially numerical implementations of the stochastic rules of motion, i.e., Langevin dynamics. In all cases, we will consider two spatial dimensions.

The paper is organized as follows. Section \ref{General} brings our review and further development of general theoretical aspects that will be used throughout the paper. In Section \ref{cons} we consider three distinct swimming strategies with constant tumbling rate. Section \ref{stoc} brings a thorough analysis of the case with stochastic tumbling rate. Our conclusions and a discussion are presented in Section \ref{conc}. Finally, the appendix~\ref{simu} gives technical details about the simulations.

\section{General theoretical aspects}
\label{General}
We start by presenting commonly used model-independent expressions which will be essential in the following sections. From these results we will then derive a general framework to more clearly extract how slowly the diffusive regime is approached. Consider a single bacterium, initially located at the origin with random orientation and internal state. The object of study is $\rho(\mathbf{r},t)$, the bacterial density at vector position $\mathbf{r}$ at time $t$ obtained by averaging over different realizations and initial states. For this initial condition [$\rho(\mathbf{r},0)=\delta(\mathbf{r})$] the bacterial density is called the van Hove function \cite{boon1991molecular}.
The MSD is
\begin{equation}
\langle r^2(t)\rangle=\int\mathrm{d}\mathbf{r} \, r^2\rho(\mathbf{r},t).
\end{equation}
When at long times the diffusive regime is achieved, the density obeys 
\begin{equation}
\frac{\partial \rho}{\partial t}=D\nabla^2 \rho,
\end{equation}
where $D$ is the diffusion coefficient, with solution in two spatial dimensions
\begin{equation}
\rho(\mathbf{r},t)=\frac{1}{4\pi D t} e^{-r^2/4Dt}. \label{eq.soldiffusion}
\end{equation}
Equation~\eqref{eq.soldiffusion} implies that $\langle r^2(t)\rangle\sim t$ and the diffusion coefficient is obtained with Einstein's relation \cite{Mazo2002}, 
\begin{equation}
D=\lim_{t\to\infty} \frac{\langle r^2(t)\rangle}{4t}. \label{DEinstein}
\end{equation}

Calculations become easier to perform through the definition of
\begin{equation}
	\tilde{\rho}(\mathbf{k},s)\equiv\int_{0}^{\infty}\mathrm{d} t \,\,  e^{-st} \int \mathrm{d}\mathbf{r} \,\,e^{-i\mathbf{k}\cdot\mathbf{r}}\rho(\mathbf{r},t)
\end{equation}
as the Laplace--Fourier transform of $\rho(\mathbf{r},t)$, where $\mathbf{k}$ is the Fourier wave vector and $s$ is the Laplace complex variable. 
Similarly to what is derived in Ref.\ \cite{boon1991molecular}, the second spatial moment (MSD) and the fourth spatial moment in 2D can be calculated, respectively, from
\begin{equation}
\langle r^2(t) \rangle=\mathcal{L}^{-1} \Bigg\{\!\!\left.-2\frac{\partial^2}{\partial k^2}\tilde{\rho}(\mathbf{k},s)\right|_{k = 0}\!\Bigg\}
\label{MSDLaplaceTrans}
\end{equation}
and
\begin{equation}
\langle r^4 (t)\rangle=\mathcal{L}^{-1} \Bigg\{\!\!\left.\frac{8}{3}\frac{\partial^4}{\partial k^4}\tilde{\rho}(\mathbf{k},s)\right|_{k = 0}\!\Bigg\},
\end{equation}
where $\mathcal{L}^{-1}$ denotes the inverse Laplace transform operator used to bring the result back to the time $t$ domain. The corresponding  long-time  diffusion coefficient $D$ can be expressed as \cite{boon1991molecular}
\begin{equation}
D=\lim\limits_{\omega\to 0} \lim\limits_{k\to 0} \frac{\omega^2}{k^2}\mathrm{Re}\left[\tilde{\rho}(\mathbf{k},i\omega)\right],
\label{LimForD}
\end{equation}
where $\omega$ is real, $k\equiv|\mathbf{k}|$, and $\mathrm{Re}(\tilde{\rho})$ denotes the real part of $\tilde{\rho}$. 

In the diffusive regime, not only the MSD must grow linearly, but also the displacement distribution must be Gaussian. In order to measure the non-Gaussianity of a particle's displacement distribution, i.e., a departure from the diffusive regime, we will be interested in the excess kurtosis, defined in 2D by
\begin{equation}
	\gamma(t)\equiv\frac{\langle r^4 \rangle}{\langle r^2 \rangle^2}-2.
\label{KurtDef}
\end{equation}
The excess kurtosis is dimensionless and vanishes for a Gaussian distribution of displacements $r$. For isotropic distributions, negative values of $\gamma$ indicate that  the distribution decays faster than a Gaussian for large displacements, while positive values implies that the distribution presents heavy tails. Notice that for one and three dimensions, one would need to subtract 3 and 5/3, respectively, instead of 2 in Eq.~\eqref{KurtDef}. At short times, when the motion is ballistic, $\langle r^4 \rangle$ equals $\langle r^2 \rangle^2$ and, therefore, $\gamma(t\to0)=-1$, as it is verified for all models presented in the next sections.
 
At instances where we give explicit expressions for the second and fourth moments, we will omit similar expressions for the excess kurtosis since they are lengthy and provide no further information. Nevertheless, expressions for their limits as well as their plots will be given and discussed.

\subsection{Extracting the excess kurtosis tail}
\label{tail}
In general, we will see in the next sections that $\langle r^2 \rangle$ and $\langle r^4 \rangle$ approach their asymptotic regimes with exponential and subdominant polynomial corrections. As a result, the excess kurtosis~ \eqref{KurtDef} approaches zero as
\begin{equation}
\gamma(t)\sim \sum_{n}  a_{n} t^{-\beta_n} e^{-\mu_{n} t},
\end{equation}
with particular sets of coefficients $a_n$, exponents ${\beta_n\geq0}$, and rates ${\mu_n\geq0}$ that  depend on the model under consideration.
We are looking for the slow decay modes to the diffusive regime, which can appear when $\mu_n$ and $\beta_n$ are small or zero. From the definition \eqref{KurtDef}, exponential factors can come from either the second or fourth moment. For example, for the second moment, we will see in the next sections that
\begin{equation}
\langle r^2 (t)\rangle\sim4 D t + \sum\limits_{n=0}^{\infty}c_n t^{\lambda_n} e^{-\mu_n t},
\end{equation}
which in Laplace space gives for small $s$
\begin{equation}
\langle \widetilde{r^2} (s)\rangle\sim\frac{4 D}{s^2} +  \sum\limits_{n=0}^{\infty} \frac{c_n\lambda_n !}{(s+\mu_n)^{1+\lambda_n}}
\end{equation}
and similarly for $\langle \widetilde{r^4}(s) \rangle$.
Hence, the exponents $\mu_n$ are recognized as minus the poles of $\partial^2_k\tilde{\rho}(\mathbf{k},s)|_{k = 0}$ and $\partial^4_k\tilde{\rho}(\mathbf{k},s)|_{k = 0}$, and the power exponents $\lambda_n$ are associated with pole multiplicity. The slowest decaying mode will be identified as the smallest  $\mu_n$.
For the majority of the models considered in this article, the long-time behavior of excess kurtosis can be explicitly obtained in real time.  However, for the last model, we will need to extract it from  Laplace space, as there is no closed expression for $\gamma(t)$.

\section{Constant tumbling rate}
\label{cons}
We will now examine three separate limiting cases of the well known Markovian run-and-tumble model. Consider a  particle moving in two spatial dimensions, for which tumbling occurs at a constant rate $\nu_0$. That is, the random walker moves with a constant speed $V$ along a body-axis $\mathbf{\hat{n}}=(\cos\theta,\sin\theta)$ that can change abruptly at a tumble event, suddenly decorrelating its orientation---in the case of \textit{E.~coli} the duration of the tumble is about ten times smaller than the duration of the runs \cite{berg2008coli} and so it is taken as zero here. The new random orientation is chosen with a kernel $W(\theta,\theta')$ that sets the probability that the swimmer changes between two specified orientation angles $\theta$ and $\theta^\prime$ at a tumble. We will assume that the space is isotropic, hence, the kernel only depends on the angle difference, $W(\theta,\theta')=w(\theta_{\mathrm{s}})$, where $w$ is an even periodic function and $\theta_{\mathrm{s}}\equiv\theta^\prime-\theta$  is the tumbling angle. 
 In addition to that, the model's particle is subject to thermal rotational diffusion with coefficient $D_{\mathrm{r}}$. Thus, in the meantime between two consecutive tumbles the orientation will change slowly and diffusively. The kinetic equation for the distribution function $f=f(\mathbf{r},\theta,t)$ is \cite{saintillan2010dilute,soto2016kinetic,Saintillan2018,saragosti2012modeling}
\begin{equation}
\frac{\partial f}{\partial t}+V\mathbf{\hat{n}}\cdot\nabla f=\nu_0\int_0^{2\pi}\!\!\! w(\theta-\theta^\prime)f(\mathbf{r},\theta^\prime,t)\mathrm{d}\theta^\prime-\nu_0 f+D_{\mathrm{r}}\nabla^2_{\mathbf{\hat{n}}}f, \label{KE.general}
\end{equation}
where the distribution function is normalized such that $\rho(\mathbf{r},t)=\int_{0}^{2\pi}f(\mathbf{r},\theta,t)\mathrm{d}\theta$. The kernel satisfies   $\int  w(\theta)\mathrm{d}\theta=1$, which guarantees that the density $\rho$ is conserved. We notice that some of the MSD results in this section are already present in some form in Refs.~\cite{shaebani2019transient,saintillan2010dilute,soto2016kinetic,Saintillan2018,saragosti2012modeling, taktikos2013motility}, but they will be developed here either as calibration of our methodology or to facilitate comparisons against new expressions such as for the excess kurtosis and with simulations. An entirely new discussion in which the MSD plays only a limited role is provided.

\subsection{Conventional run-and-tumble model with complete reorientation} \label{sec.classicalRT}
We start with the limiting case where $D_{\mathrm{r}}=0$ and there is complete reorientation after tumbling, that is, $w(\theta_{\mathrm{s}})=1/2\pi$, the simplest version of the run-and-tumble model. Although no known microswimmer reorients completely after a tumble event, this model will serve to calibrate our methodology, as mentioned. In this case the kinetic equation for the probability density function $f(\mathbf{r},\theta,t)$ is just
\begin{equation}
\frac{\partial f}{\partial t}+V\mathbf{\hat{n}}\cdot\nabla f=\frac{\nu_0}{2\pi}\int_{0}^{2\pi} f(\mathbf{r},\theta^\prime,t)\mathrm{d}\theta^\prime-\nu_0 f.
\label{ClassRT}
\end{equation}
The initial condition is $f(\mathbf{r},\theta,0)=\delta(\mathbf{r})/2\pi$, meaning that the initial orientation is random. With a view to obtaining $\tilde{\rho}(\mathbf{k},s)$ satisfying this kinetic equation, we move to Laplace--Fourier space. This leads to Eq.~(\ref{ClassRT}) being rewritten as
\begin{equation}
\left(s+iV\mathbf{k}\cdot\mathbf{\hat{n}}+\nu_0\right)\tilde{f}=\frac{1}{2\pi}\left(1+\nu_0\tilde{\rho}\right),
\end{equation}
where we used that 
\begin{equation}
\tilde{\rho}(\mathbf{k},s)=\int_{0}^{2\pi} \tilde{f}(\mathbf{k},\theta,s)\mathrm{d}\theta
\end{equation}
for the Laplace--Fourier transform $\tilde{f}(\mathbf{k},\theta,s)$ of the distribution function.
Therefore by isolating $\tilde{f}(\mathbf{k},\theta,s)$ and integrating over $\theta$ we obtain a closed equation for $\tilde\rho$, which gives
\begin{align}
\tilde{\rho}(\mathbf{k},s)&=\frac{1}{\sqrt{(s+\nu_0)^{2}+V^{2}k^{2}}-\nu_0}, \label{rho.RTclasico}\\
&=\frac{1}{s}-\frac{V^2 k^2}{2s^2(s+\nu_0)} + \frac{(3s+2\nu_0)V^4 k^4}{8 s^3(s+\nu_0)^3}+\mathcal{O}(k^6), 
\end{align}
where in the second line we made a Taylor expansion in $k$ to easily identify the poles associated to the second and fourth moments. 
We can now use the equations in Section \ref{General} to obtain our desired quantities. The MSD is
\begin{equation}
\langle r^{2}\rangle=\frac{2V^{2}}{\nu_0^{2}}\left(\nu_0 t+e^{-\nu_0 t}-1\right),
\end{equation}
from which one can either use Einstein's relation \eqref{DEinstein} or directly apply Eq.~(\ref{LimForD}) to obtain
\begin{equation}
D=\frac{V^2}{2\nu_0},
\label{Coeff}
\end{equation}
which is a widely known result \cite{romanczuk2012active, taktikos2011modeling,bechinger2016active}. The fourth spatial moment reads
\begin{equation}
\langle r^{4}\rangle=\frac{4 V^4}{\nu _0^4} \left[2  \left(\nu _0^2
t^2-3\nu_0 t+3\right) + e^{-\nu _0 t}(\nu _0^2 t^2-6)\right],
\end{equation}
allowing one to compute the excess kurtosis directly through \eqref{KurtDef}.
The kurtosis longest-standing exponential goes as $\exp(-\nu_0 t)$, which does not present any singular behavior.

\begin{figure}
		\centering
		\includegraphics[width=\linewidth]{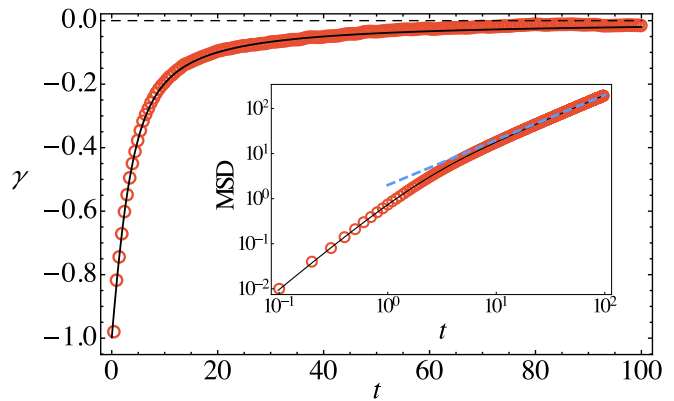}
	\caption{Conventional run-and-tumble model with complete reorientation (to calibrate our methodology): theory (solid black line) and simulation (circles) for the time evolution of the excess kurtosis $\gamma$ and, in the inset, of the MSD (log-log scale). The dashed line is $4Dt$ where the diffusion coefficient $D$ is given by Eq.~(\ref{Coeff}). Units are chosen such that $V=\nu_0=1$.}
\label{Fig1}
\end{figure}

In Fig.~\ref{Fig1} the above expressions for the MSD, the diffusion coefficient, and the excess kurtosis are tested against our simulations, which have been performed by directly solving the associated run-and-tumble motion equations (see the appendix~\ref{simu} for details on the simulation method). The agreement is excellent as expected since no approximations were made.

\subsection{Partial reorientation} \label{sec.anisotropic}
We now generalize the previous analysis to the case of partial reorientation while keeping $D_{\mathrm{r}}=0$. In this case the kernel is no longer uniformly distributed between $0$ and $2\pi$ and it  is fully characterized by its cosine Fourier components 
\begin{equation}
\sigma_{n}\equiv\langle\cos\left(n\theta_{\mathrm{s}}\right)\rangle = \int_{-\pi}^{\pi} \mathrm{d}\theta_{\mathrm{s}}\, w(\theta_{\mathrm{s}})\cos\left(n\theta_{\mathrm{s}}\right),\quad n\geq 1,
\end{equation}
which, in the previous case, vanish completely.
This model accounts for many flagellated bacteria and unicellular algae~\cite{stocker2009tumbling}. For the case of \textit{E. coli}, the kernel has been measured~\cite{berg1972chemotaxis}, giving $\sigma_1\simeq0.33$~\cite{lovely1975statistical}.
 It can be shown that the mean-squared displacement depends on $\sigma_1$ only~\cite{taktikos2013motility}. Thus, for the purpose of computing this quantity, only the average value $\sigma_1$ matters and so we do not need to worry about the whole shape of $w$. 
 However, we show below that the excess kurtosis and the crossover time to reach the diffusive regime depend also on $\sigma_2$.

The Laplace--Fourier transform of the kinetic equation~\eqref{KE.general} for this case is
\begin{equation}
\left(s+iV\mathbf{k}\cdot\mathbf{\hat{n}}+\nu_0\right)\tilde{f}=\frac{1}{2\pi}+\nu_0\int_0^{2\pi} \!w(\theta-\theta^\prime)\tilde{f}(\mathbf{k},\theta^\prime,s)\mathrm{d}\theta^\prime,
\end{equation}
where we used the same initial condition as in Sec.~\ref{sec.classicalRT}. 
To solve it, we expand the distribution function in Fourier modes
\begin{equation}
\tilde{f}(\mathbf{k},\theta,s)=\sum_{n=0}^{\infty}\left[h_{n}\cos(n\theta)+g_{n}\sin(n\theta)\right], \label{fourierseriesf}
\end{equation}
where the coefficients $h_{n}$ and $g_{n}$ depend on $\mathbf{k}$ and $s$, and are to be determined by plugging the solution into the kinetic equation. 
Taking $\mathbf{k}=k\mathbf{\hat{x}}$, it is clear that the sine modes will vanish identically, and so we can set $g_n=0$ from now on.
The convolution integral can be expressed as
\begin{equation}
\int_0^{2\pi} w(\theta-\theta^\prime)\tilde{f}(\mathbf{k},\theta^\prime,s)\mathrm{d}\theta^\prime=\sum_{n=0}^{\infty}\sigma_n h_{n}\cos(n\theta).
\end{equation}
By keeping terms up to $n=2$, we truncate the Fourier series, which allows us to obtain a closed expression for $\tilde{\rho}(\mathbf{k},s)$. 
For the sake of presentation the long result is expressed as an expansion up to fourth order in $k$. This has no implications as no higher-order derivative in $k$ will be required. We have
\begin{multline}
\tilde{\rho}(\mathbf{k},s)=\frac{1}{s}-\frac{V^2k^2 }{2 s^2 \left(\nu _0(1-\sigma_1)+s\right)}\\
+\frac{\left(3 s-2 \nu _0 \left(\sigma _2-1\right)\right) V^4k^4}{8 s^3 \left(\nu _0(1-\sigma_1)+s\right){}^2 \left(\nu
	_0(1-\sigma _2)+s\right)} + \mathcal{O}(k^6). \label{rhoks.aniso}
\end{multline}

Using the expressions of Sec.~\ref{General}, the MSD is
\begin{equation}
\langle r^{2}\rangle=\frac{2V^{2}}{\nu_0^{2}(1-\sigma_1)}\left[\nu_0  t +\frac{e^{-\nu_0 (1-\sigma_1)t}-1}{(1-\sigma_1)}\right],
\label{MSDpartial}
\end{equation}
with diffusion coefficient
\begin{equation}
D=\frac{V^2}{2\nu_0(1-\sigma_1)},
\end{equation}
which is a well known result~\cite{soto2016kinetic,saragosti2012modeling}. The fourth moment is
\begin{multline}
\langle r^{4}\rangle=\frac{8 V^4}{\nu _0^4} \bigg[
\frac{\nu _0^2 t^2}{\left(1-\sigma _1\right){}^2} 
+\frac{e^{-\nu _0 \left(1-\sigma
			_2\right) t}}{\left(\sigma _1-\sigma _2\right){}^2 \left(1-\sigma _2\right){}^2}\\
-\frac{\sigma _1^2+2 \left(\sigma _2-2\right) \sigma _1-6 \sigma _2^2+10 \sigma _2-3}{\left(1-\sigma _1\right){}^4 \left(1-\sigma
		_2\right){}^2}	
				\\
		-\frac{\nu _0 \left(3 \sigma _1-2 \sigma _2-1\right) t e^{-\nu _0 \left(1-\sigma
				_1\right) t}}{\left(1-\sigma _1\right){}^3 \left(\sigma _1-\sigma _2\right)}+\frac{\nu _0 \left(\sigma _1-4 \sigma _2+3\right) t}{\left(1-\sigma
		_1\right){}^3 \left(\sigma _2-1\right)}
	\\
		-\frac{\left(9 \sigma _1^2-2 \left(7 \sigma _2+2\right) \sigma _1+6 \sigma _2^2+2 \sigma _2+1\right)
		e^{-\nu _0 \left(1-\sigma _1\right) t}}{\left(1-\sigma _1\right){}^4 \left(\sigma _1-\sigma _2\right){}^2}
		\bigg]. \label{r4.anisotropic}
\end{multline}

While for the complete-reorientation kernel of Sec.~\ref{sec.classicalRT} there is a single relaxation time, for a general kernel two relaxation rates appear: $\nu_1=\nu_0(1-\sigma_1)$ and $\nu_2=\nu_0(1-\sigma_2)$. In this regard, the complete-reorientation case is singular since the two relaxation times merge, increasing the multiplicity of the corresponding pole in \eqref{rhoks.aniso}. This implies that, while for the complete-reorientation case the excess kurtosis decays purely exponentially as $\gamma\sim\exp(-\nu_0 t)$, here $\gamma$ is the sum of  two leading terms, $\exp(-\nu_1 t)/t$ and $\exp(-\nu_2 t)/t^2$, except for the singular case where both rates are equal, in which ${\gamma\sim\exp(-\nu_{1,2} t)}$. 

\begin{figure}[htb]
	\centering
	\includegraphics[width=\linewidth]{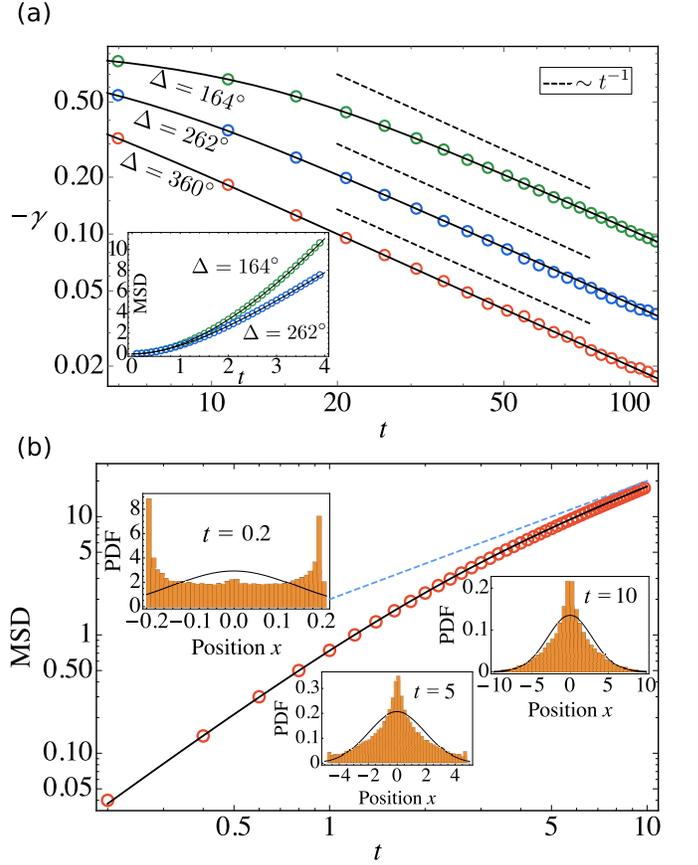}
	\caption{Run-and-tumble model with partial reorientation. (a) Theory (solid black lines) and simulation (circles) for (the negative of) the excess kurtosis as a function of time in log-log scale. The MSDs are shown in the inset in linear scale to highlight the departure between the parameters.  The values of $\Delta$ are indicated in degrees: $164^{\circ}$, $262^{\circ}$, and, the complete-reorientation limit, $360^{\circ}$ (green, blue, and red, respectively). 
		(b) MSD in log-log scale for a kernel uniformly distributed around both $0^{\circ}$ and $180^{\circ}$ with $\Delta=20^{\circ}$ as defined in the main text, giving $\sigma_1=0$ and $\sigma_2\simeq0.99$. Theory (solid black line), simulation (circles), and the linear part of the MSD (dashed blue line).
		Insets: probability distribution function of the $x$-displacement at different time instants as from simulations (solid lines are normalized Gaussian distributions with the same mean and variance as the corresponding data). The excess kurtosis (not shown) changes from negative to positive at $t\simeq4.76$. Units are chosen such that $V=\nu_0=1$. }
	\label{Fig2}
\end{figure}

The approach to a linear time dependence in the MSD is controlled by the relaxation time $T_1=1/\nu_1$, which diverges when the average tumbling angle is small. Naturally, in this case, when swimmers deviate little in each tumble event the persistence is enhanced, implying a large diffusion coefficient. Importantly, also the  amplitude of the non-diffusive term diverges when $\sigma_1\approx 1$, making such a departure from diffusion more relevant. 
Figure \ref{Fig2}a shows this behavior. To compare with simulations, first we consider the case in which the tumbling angles  are uniformly distributed in the range $[-\Delta/2,\Delta/2]$. 
The second relaxation time, $T_2=1/\nu_2$, appears in the fourth moment given by Eq.~\eqref{r4.anisotropic}. Both the relaxation time and the associated amplitude diverge when $\sigma_2\approx 1$, implying that for long times the displacement distribution deviates largely from a Gaussian one. 
Together with the slow exponential decay, algebraic terms also contribute to $\gamma$ with amplitudes that can be quite large as they read $(2T_2-4T_1)/t$. 
It can be seen that the smaller the $\Delta$ the slower is the excess kurtosis approach to zero.
Note that, for this kernel, the two relaxation times scale as $1/(\nu_0\Delta^2)$ and are of a similar order, implying that both conditions for the diffusive regime to be valid---the linear increase of the MSD and a small excess kurtosis---are attained in the same timescale.
 
It is possible, however, that $T_1$ and $T_2$ decouple if $\sigma_2\approx 1$ and, simultaneously, $\sigma_1$ is far from $1$. Then although the MSD reaches the linear regime rapidly, the excess kurtosis remains finite and positive for long times, implying that the diffusion equation is \textit{not} valid in this period. This situation occurs, for example, if the tumbling angles distribution is sharply centered around both 0 and $180^{\circ}$.
Figure 2b shows the case where $\theta_{\mathrm{s}}$ is uniformly distributed in the ranges $[-\Delta/4,\Delta/4]$ and $[180^{\circ}-\Delta/4,180^{\circ}+\Delta/4]$, that is, any new tumbling angle $\theta_s$ is randomly drawn out of these two ranges. In this case, a small $\Delta$ leads indeed to a sharp separation of the time scales $T_1$ and $T_2$. As a result, we can see in Fig.~\ref{Fig2}b that the MSD becomes linear in time even if the displacement distribution is still strongly non-Gaussian, as revealed by the insets.
For this class of kernels, tumbling gives rise for a single swimmer  to a one-dimensional random walk along $\mathbf{\hat{n}}$ and only slowly, with a rate proportional to the dispersion of tumbling angles around 0 and $180^{\circ}$, i.e., $\sigma_2$, the process evolves to a two-dimensional  diffusion. For a collection of swimmers initially seeded at $\mathbf{r}=0$, the intermediate dynamics for $T_1<t<T_2$ will therefore be diffusive only in the radial direction.  

Also, our analytical results for the MSD in Eq.~\eqref{MSDpartial} and for the excess kurtosis [from Eq.~\eqref{KurtDef} using \eqref{MSDpartial} and \eqref{r4.anisotropic}] are compared against the simulations in Fig.~\ref{Fig2}. We highlight that they agree well with simulations (even for small values of $\Delta$, not shown) despite the approximation made in truncating the Fourier series up to $n=2$. 

\subsection{Run-and-reverse with thermal rotational diffusion}
\begin{figure}[!ht]
		\centering
		\includegraphics[width=\linewidth]{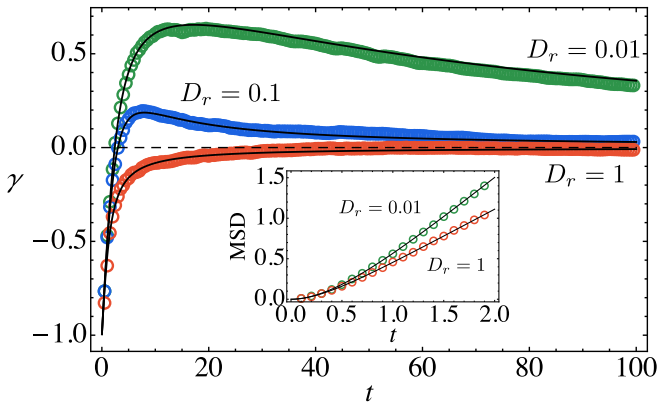}
	\caption{Run-and-reverse model with rotational diffusion: theory (solid black lines) and simulation (circles) for the excess kurtosis $\gamma$ as a function of time (and for the MSD in the inset). The values of $D_{\mathrm{r}}$ are indicated: $0.01$, $0.1$, and $1$ (green, blue, and red, respectively). Units are chosen such that $V=\nu_0=1$. The MSDs are shown in linear scale to highlight the departure between the parameters.}
	\label{Fig3}
\end{figure}

As already mentioned, up to $70\%$ of marine bacteria are believed to have a distribution of tumbling angles peaked around $180^{\circ}$ \cite{johansen2002variability}. The soil bacteria \textit{Bradyrhizobium diazoefficiens} has also been shown to perform this kind of tumbling~\cite{quelas2016swimming}.
 In the limiting case known as run-and-reverse dynamics, which we consider now, the particle's tumble can only lead to the exactly opposite motion direction. In this limit it becomes physically unreasonable to neglect thermal diffusion and so we will take $D_{\mathrm{r}}>0$; otherwise the swimmer will indefinitely perform a one-dimensional random walk. In this model, $w(\theta_{\mathrm{s}})=\delta(\theta_{\mathrm{s}}-\pi)$, and hence the kinetic equation reads
\begin{equation}
\frac{\partial f}{\partial t}+V\mathbf{\hat{n}}\cdot\nabla f=\nu_0f(\mathbf{r},\theta+\pi,t)-\nu_0 f+D_{\mathrm{r}}\frac{\partial^{2} f}{\partial\theta^{2}},
\end{equation}
where we notice that the indicated instance of $f$ is evaluated at $\theta+\pi$, while the other ones are evaluated at $\theta$ as per usual. 
After the Laplace--Fourier transform is applied and using  the same initial condition as before, i.e., $f(\mathbf{r},\theta,0)=\delta(\mathbf{r})/2\pi$,  we obtain
\begin{equation}
\left(s+iV\mathbf{k}\cdot\mathbf{\hat{n}}+\nu_0\right)\tilde{f}-\nu_0\tilde{f}(\mathbf{k},\theta+\pi,s)-D_{\mathrm{r}}\frac{\partial^2 \tilde{f}}{\partial\theta^2}=\frac{1}{2\pi}.
\end{equation}
As in the previous case, we expand $\tilde{f}$ in a Fourier series [Eq.~\eqref{fourierseriesf}], where again $g_n=0$ by symmetry.
We truncate the series keeping only the terms $n\leq2$ and solve for the coefficients. Integrating $\tilde{f}(\mathbf{k},\theta,s)$ over $\theta$, we obtain
\begin{widetext}
\begin{equation}
\tilde{\rho}(\mathbf{k},s)=
\frac{4 \left(4 D_{\mathrm{r}}+s\right) \left(D_{\mathrm{r}}+2 \nu _0+s\right)+V^2 k^2 }{8 V^2 k^2  D_{\mathrm{r}}+20 D_{\mathrm{r}} s^2+8 \nu _0 s \left(4 D_{\mathrm{r}}+s\right)+16 s D_{\mathrm{r}}^2+3s
	V^2 k^2 +4 s^3}. \label{rhoks.randr}
\end{equation}

Upon using the formulae in Section \ref{General}, we find that the MSD is given by
\begin{equation}
\langle r^{2}\rangle=\frac{2V^{2}}{(D_{\mathrm{r}}+2\nu_0)^{2}}\left[(D_{\mathrm{r}}+2\nu_0)t+e^{-(D_{\mathrm{r}}+2\nu_0)t}-1\right],
\label{MSDRR}
\end{equation}
with diffusion coefficient 
\begin{equation}
D=\frac{V^{2}}{2(D_{\mathrm{r}}+2\nu_0)},
\end{equation}
while the fourth moment is\\
\begin{equation}
\begin{aligned}
\langle r^{4}\rangle={} &\frac{V^4}{2} \left[\frac{87 D_{\mathrm{r}}^2-4 \nu _0^2-20 \nu _0 D_{\mathrm{r}}}{D_{\mathrm{r}}^2 \left(D_{\mathrm{r}}+2 \nu _0\right){}^4}+
\frac{16 t^2}{\left(D_{\mathrm{r}}+2 \nu
	_0\right){}^2}+\frac{8 \nu _0 t-60 D_{\mathrm{r}} t}{D_{\mathrm{r}} \left(D_{\mathrm{r}}+2 \nu _0\right){}^3}+\frac{e^{-4 D_{\mathrm{r}} t}}{D_{\mathrm{r}}^2 \left(3 D_{\mathrm{r}}-2 \nu _0\right){}^2}\right. \\[20pt]
&\left.-\frac{16 e^{-\left(D_{\mathrm{r}}+2 \nu _0\right) t} \left(D_{\mathrm{r}}^2
	\left(2 \nu _0 t+49\right)-4 \nu _0 D_{\mathrm{r}} \left(11 \nu _0 t+19\right)+15 D_{\mathrm{r}}^3 t+12 \nu _0^2 \left(2 \nu _0 t+3\right)\right)}{\left(3
	D_{\mathrm{r}}-2 \nu _0\right){}^2 \left(D_{\mathrm{r}}+2 \nu _0\right){}^4}\right].
\end{aligned}
\label{M4DRR}
\end{equation}
\end{widetext}
The MSD can rapidly reach a regime where it grows linearly with time, but for small rotational diffusion, the process remains non-Gaussian, with large positive values of $\gamma$. Similarly to the previous case, when the scattering angle is narrowly distributed around $0^{\circ}$ and $180^{\circ}$, reversions at rate $\nu_0$ induce a one-dimensional diffusive motion along the director axis, but an authentic two-dimensional diffusion is only achieved at a typical time $1/(4D_{\mathrm{r}})$ as the axis changes direction. For a perfect one-dimensional random walk, i.e. for $D_r=0$, the excess kurtosis equals 1 as $\gamma=\langle x^4\rangle/\langle x^2 \rangle^2-2=3-2$. For small enough rotational diffusion, the excess kurtosis becomes positive, having a peak that can be quite large (see Fig.~\ref{Fig3}), reflecting this quasi one-dimensional motion. Positive excess kurtosis with a very slow decay also appear in the similar case of partial reorientation without rotational diffusion (Sec.~\ref{sec.anisotropic}) for $\sigma_2\approx1$, as can be seen directly from Eq.~\eqref{r4.anisotropic}.

\section{Stochastic tumbling rate}
\label{stoc}
In bacteria like \textit{E.~coli}, the tumbling process is triggered by a reversion in the sense of rotation (from counter-clockwise, CCW, to clockwise, CW) of one or several flagella. As a result, the flagella bundle dissembles and the propulsion thrust is lost \cite{chen2000torque}. By analyzing the biochemistry of the molecular motor, Tu and Grinstein proposed that the tumbling process can be described as a two state activated system, where the free energy barrier to transit from the CCW to the CW state depends sensibly on the concentration inside the bacterial body of the so-called CheY-P protein, denoted by $[Y]$~\cite{tu2005white}. In the Tu--Grinstein model the tumble rate is $\nu=\bar{\nu} \exp(-G([Y])/k_\mathrm{B}T)$, where $G$ is the free energy barrier and $\bar{\nu}$ a constant. Expanding $G$ around the average value $[Y_0]$, they propose
\begin{equation}
\nu(X) = \nu_0 e^{\alpha X}, \label{nuofX}
\end{equation}
where $X(t)=([Y](t)-[Y_0])/\sigma_Y$ corresponds to the fluctuations in concentration normalized to $\sigma_Y$, the standard deviation of $[Y]$. Finally,  $\nu_0$ absorbs all the prefactors. Note that $\nu_0$ has been used in the previous sections to denote the tumbling rate of models without stochasticity, that is, where the tumbling rate is constant over time. Here we use it with exactly the same meaning: in the limit where $\alpha\rightarrow0$ the tumbling rate is $\nu(X)\rightarrow\nu_0$. The parameter $\alpha$ is positive \cite{cluzel2000ultrasensitive} and quantifies the sensitivity of the system to changes in the protein concentration. 
This phosphorylated protein has a small production rate, with a long memory time $T$, and consequently $X$ is well described by the Ornstein-Uhlenbeck process
\begin{equation}
\frac{\mathrm{d}X}{\mathrm{d}t}=-\frac{X}{T}+\sqrt{\frac{2}{T}}\xi(t),
\label{Xevol}
\end{equation}
where $\xi$ is an additive zero-mean Gaussian white noise with correlation $\langle \xi(t)\xi(t')\rangle=\delta(t-t')$. By tracking several individual \textit{E.~coli} bacteria it has been possible to fit the model parameters to $T=\SI{19.0}{\second}$, $\nu_0 = \SI{0.65}{\second^{-1}}$, and 
$\alpha = 1.62$~\cite{figueroa20183d}.
The same experiments gave for the rotational diffusivity $D_{\mathrm{r}}=\SI{0.025}{\second^{-1}}$ and for the tumbling $\sigma_1=0.112$. Considering that $D_{\mathrm{r}}\ll\nu_0$ and that $\sigma_1\approx 0$, we will consider complete reorientation after tumbling and neglect the rotational diffusion. This approximation also helps to highlight the new phenomenology that appears from considering the internal variable $X$.

With $X$ as a new variable of the distribution function, the kinetic equation for $f=f(\mathbf{r},\theta,X,t)$ reads
\begin{multline}
\frac{\partial f}{\partial t}+V\mathbf{\hat{n}}\cdot\nabla f=
\frac{1}{T}\left[\frac{\partial^{2} f}{\partial X^{2}}+\frac{\partial (Xf)}{\partial X}\right]\\+
\frac{\nu(X)}{2\pi}\int_{0}^{2\pi} f(\mathbf{r},\theta',X,t)\mathrm{d}\theta^\prime-\nu(X)f.
\label{StocTumbEqn}
\end{multline}
where the distribution function is normalized such that $\rho(\mathbf{r},t) = \int f(\mathbf{r},\theta,X,t) \mathrm{d}\theta \mathrm{d}X $.

Once again, we change to the Laplace--Fourier space and so Eq.~(\ref{StocTumbEqn}) becomes
\begin{multline}
s\tilde{f}-\frac{1}{(2\pi)^{3/2}}e^{-X^{2}/2}+iV\mathbf{k}\cdot{\mathbf{\hat{n}}}\tilde{f}=\\
\frac{1}{T}\left[\frac{\partial^{2}\tilde{f}}{\partial X^{2}}+\frac{\partial (X\tilde{f})}{\partial X}\right]+\nu(X)\left[\frac{\tilde{g}(\mathbf{k},X,s)}{2\pi}-\tilde{f}\right],
\label{TransStocTumbEqn}
\end{multline}
where $\tilde f$ stands for $\tilde{f}(\mathbf{k},\theta,X,s)$, and we have made use of the definition
\begin{equation}
\tilde{g}(\mathbf{k},X,s)\equiv\int_{0}^{2\pi}\tilde{f}(\mathbf{k},\theta^\prime,X,s)\mathrm{d}\theta^\prime
\end{equation}
and the initial condition
\begin{equation}
f(\mathbf{r},\theta,X,t=0)=\frac{1}{(2\pi)^{3/2}}e^{-X^{2}/2}\delta(\mathbf{r}),
\label{varxIC}
\end{equation}
which indicates that the internal variable $X$ is in equilibrium.  
We propose the solution
\begin{equation}
\tilde{f}(\mathbf{k},\theta,X,s)=\sum_{n=0}^{\infty} G_{n}(X)\tilde{f}_{n}(\mathbf{k},\theta,s),
\label{sol}
\end{equation}
where the coefficients $\tilde{f}_{n}(\mathbf{k},\theta,s)$ do not depend on $X$ and
\begin{equation}
G_{n}(X)\equiv e^{-X^{2}/2}H_{n}(X/\sqrt{2}),
\end{equation}
in which $H_{n}$ is the Hermite polynomial of order $n$ [such that $H_0(x)=1$, $H_1(x)=2x$, \ldots]~\cite{arfken1999mathematical}. Using the eigenvalue equation for the Hermite polynomials allows us to write 
\begin{multline}
\frac{1}{T}\left[\frac{\partial^{2}\tilde{f}(\mathbf{k},\theta,X,s)}{\partial X^{2}}+\frac{\partial (X\tilde{f}(\mathbf{k},\theta,X,s))}{\partial X}\right]=\\
-\frac{1}{T}\sum_{n=0}^{\infty}{n} e^{-X^{2}/2}H_{n}(X/\sqrt{2})\tilde{f}_{n}(\mathbf{k},\theta,s).
\label{1}
\end{multline}
Since our goal is to find the Laplace--Fourier transform of $\rho(\mathbf{r},t)$, i.e., $\tilde{\rho}(\mathbf{k},s)$, which does not depend on $\theta$, it is helpful to define $\tilde{g}_{n}(\mathbf{k},s)=\int_{0}^{2\pi}\tilde{f}_{n}(\mathbf{k},\theta,s)\mathrm{d}\theta$.
 
 At this point we proceed by plugging the  above equations into Eq.~(\ref{TransStocTumbEqn}), then multiplying by $H_m(X/\sqrt{2})$, and finally integrating over $X$.
 One obtains
\begin{equation}
\sum_{n=0}^{\infty}A_{mn}(\mathbf{k},\theta,s)\tilde{f}_{n}(\mathbf{k},\theta,s)=c_{m}+\sum_{n=0}^{\infty}B_{mn}\tilde{g}_{n}(\mathbf{k},s),
\label{MatrixEqn}
\end{equation}
where
\begin{equation}
	A_{mn}(\mathbf{k},\theta,s)\equiv 
	2^{n}n!\sqrt{\pi}\delta_{mn} \left(s+iV\mathbf{k}\!\cdot\!\mathbf{\hat{n}}+\frac{{n}}{T}\right)+\nu_0 J_{mn},
\end{equation}
\begin{equation}
B_{mn}\equiv\frac{\nu_{0}}{2\pi}J_{mn}, \quad c_{m}\equiv\frac{1}{2\pi\sqrt{2}}\delta_{m0},
\end{equation}
where the $\delta_{ij}$ are Kronecker deltas and 
\begin{equation}
J_{mn}\equiv\int_{-\infty}^{\infty}e^{-y^{2}}e^{\sqrt{2}\alpha y}H_{n}(y)H_{m}(y)\mathrm{d}y.
\end{equation}

The linear Eqs.~\eqref{MatrixEqn} can be solved for $\tilde{f}_n$ in terms of $\tilde{g}_n$. Integrating over $\theta$ gives now a closed linear set of equations for $\tilde{g}_n$, which can be directly solved. Noting that $\tilde{\rho}=\tilde{g}_0$ (which can be seen through the orthogonality between $H_0$ and $H_n$), one obtains
\begin{multline}
\tilde{\rho}(\mathbf{k},s)=\int_{0}^{2\pi} \bigg[\frac{1}{2\sqrt{\pi}}(\mathbf{A}^{-1})_{00}(\mathbf{k},\theta,s)  \\
+ \sqrt{2\pi} \sum_{m,n=0}^{\infty} {(\mathbf{A}^{-1})_{0m}}(\mathbf{k},\theta,s) B_{mn}\tilde{g}_{n}(\mathbf{k},s)\bigg] \mathrm{d}\theta.
\label{FinalRhoStocCase}
\end{multline}

To obtain explicit expressions, Eq.~(\ref{FinalRhoStocCase}) is truncated at a certain order $n=m=N_{\rm max}$. The greater the $N_{\rm max}$ the higher is the order of a polynomial in $\alpha$ that appears in $J_{mn}$. Hence, increasing $N_{\rm max}$ one increases the range in $\alpha$ over which the theory is valid. However, the greater the $N_{\rm max}$ the more complicated are the elements of the inverse of $\mathbf{A}$, which eventually need to be integrated in $\theta$. Therefore $N_{\rm max}$ also affects how complicated it is the $\tilde{\rho}(\mathbf{k},s)$ over which one needs to apply the inverse Laplace transform as well as to compute limits. As it turns out, those complications grow rapidly with $N_{\rm max}$, with the case $N_{\rm max}=0$ being the only one that we have treated fully analytically. The $\tilde{\rho}(\mathbf{k},s)$ obtained by expanding up to this order is identical to the conventional RT case \eqref{rho.RTclasico}, provided that one considers the tumbling rate to be $\nu_0 \exp{(\alpha^2/2)}$, which corresponds to the average of Eq.~\eqref{nuofX} over $X$. 
See Section \ref{Tlim} for the related analysis of the limits $T\to0$ and $T\to\infty$.

\begin{figure}[!ht]
	\centering
	\includegraphics[width=\linewidth]{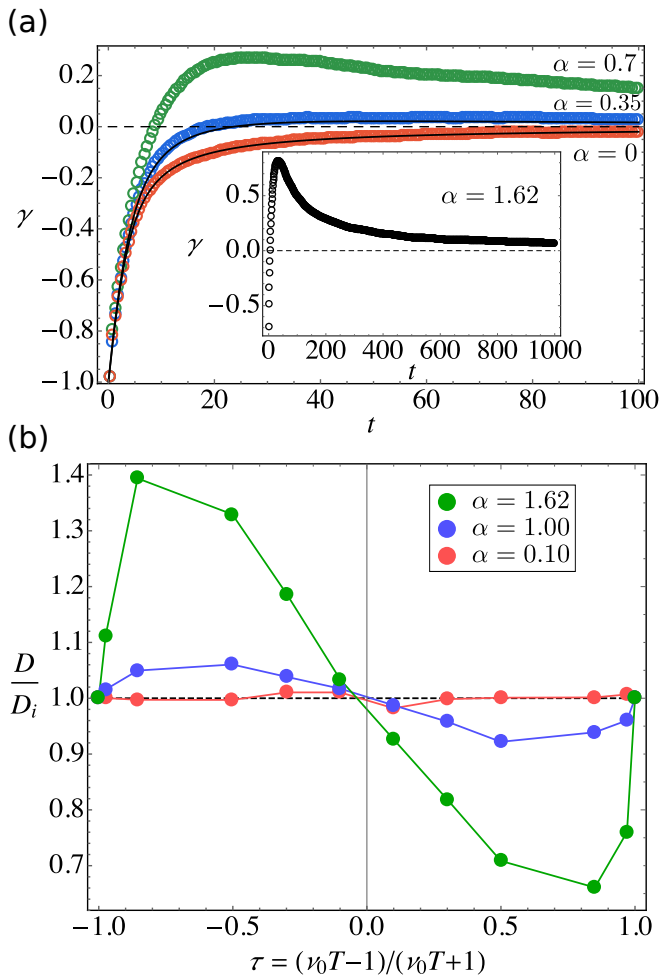}
	\caption{Run-and-tumble model with stochastic tumbling rate. 
	(a) Excess kurtosis as a function of time for $\nu_0 T=12.3$ and different values of $\alpha$. The case $\alpha=1.62$, which corresponds to the experimentally fitted value for \textit{E. coli} is shown in the inset as it falls out of scale. The circles are the simulation results while the solid lines are the small-$\alpha$ theoretical prediction (only for $\alpha=0$ and $\alpha=0.35$).
	(b) Diffusion coefficient $D$ scaled by the interpolating expression \eqref{DmodelX.interpolated} as a function of $\tau=(\nu_0 T-1)/(\nu_0 T+1)$ for different values of $\alpha$. The points at $\tau=\pm1$ are not obtained from simulations but rather from the asymptotic expressions for the zero- and infinite-memory limits. Units are chosen such that $V=\nu_0=1$.}
	\label{Fig4}
\end{figure}

For $N_{\rm max}=1$ new physics is found. Although involved, it is possible to obtain an explicit expression for $\tilde{\rho}(\mathbf{k},s)$ from where the diffusion coefficient is obtained using Eq.~\eqref{LimForD},
\begin{equation}
D=\frac{V^2 }{2 \nu_0 e^{\alpha^2/2}}\left(1 + \frac{\alpha^2 \nu_0  T}{1+\nu_0  T}\right). \label{D.smallalpha}
\end{equation}
It is not possible, however, to analytically perform the inverse transforms of the second and fourth moments. Instead, they are calculated by applying a semi-numerical inverse Laplace transform method for comparison with simulations. As one can see in Fig.~\ref{Fig4}, the analytical results in this case agree very well with simulations up to a significant value of $\alpha$. The higher the $\alpha$, the higher the peak in the excess kurtosis.

Despite the aforementioned complications in obtaining an expression for $\tilde{\rho}(\mathbf{k},s)$, we can still use the method in Section \ref{tail} to extract the late-time exponential decay of the excess kurtosis. The second and fourth moments share poles, and upon reversing their sign we obtain
\begin{align}
\nu_1&=\nu_0[1+\alpha^2(1/2-T\nu_0) + \mathcal{O}(\alpha^4)],\\
\nu_2&= \nu_0[1+1/(T\nu_0)+ \alpha^2 (3/2 + T \nu_0)+ \mathcal{O}(\alpha^4)].
\end{align}
While the second rate remains finite for all values of the parameters, $\nu_1$  decays linearly with $T$. 
Therefore the greater the protein memory the longer it will take for diffusion to be achieved. But the behavior for varying $\alpha$ depends on where in the memory $T$ range we are: if $T\nu_0>1/2$ then $\nu_1$ also decays with $\alpha$, meaning a slower approach to diffusion, and if $T \nu_0$ is smaller than that then increasing $\alpha$ speeds up the approach. To evaluate the importance of this eventual slow approach to diffusion, we compute the multiplicity and amplitude of the associated pole, obtaining
\begin{equation}
\langle \widetilde{r^4}(s)\rangle\sim \frac{8 V^4[1+\nu_0 T(1+2\nu_0 T)\alpha^2/2+\mathcal{O}(\alpha^4)]}{\nu_0^2(s+\nu_1)^3}
\end{equation}
This implies that at long times the excess kurtosis exponential decay is $\gamma\sim\exp(-\nu_1 t)$, with an amplitude that grows with $\alpha$, in agreement with the results shown in Fig.~\ref{Fig4}. 

\subsection{Zero- and infinite-memory limits}
\label{Tlim}
In the limiting case of very small memory time $T$, $X$ fluctuates rapidly and the tumble rate is effectively an average of \eqref{nuofX} over all possible values of $X$, that is, $\langle\nu\rangle= \nu_0 \exp{(\alpha^2/2)}$. This result can be achieved more formally by expanding the distribution function $\tilde{f}$ for small $T$ as $\tilde{f}=\tilde{f_0}+T\tilde{f_1}+\mathcal{O}(T^2)$ and $\tilde{g}=\tilde{g_0}+T\tilde{g_1}+\mathcal{O}(T^2)$, and replacing these into the  Laplace--Fourier-transformed kinetic equation (\ref{TransStocTumbEqn}). For $\mathcal{O}(1/T)$ we obtain a simple differential equation in $X$ for $\tilde{f_0}$ whose solution can be cast as $\tilde{f_0}=e^{-X^2/2}a(\mathbf{k},\theta,s)$ where $a(\mathbf{k},\theta,s)$ is some coefficient function independent of $X$. 
At $\mathcal{O}(T)$ the equation reads
\begin{multline}
	iV\mathbf{k}\!\cdot\!\mathbf{\hat{n}}\, e^{-X^2/2}\,a+s\, e^{-X^2/2}\,a+\nu _0 e^{-X^2/2+\alpha  X}\,a\\
			-\frac{\nu _0 e^{-X^2/2+\alpha  X}\,b}{\sqrt{2 \pi }}-\frac{e^{-X^2/2}}{(2\pi) ^{3/2}}=\frac{\partial^{2} \tilde{f}_1}{\partial X^{2}}+\frac{\partial (X \tilde{f}_1)}{\partial X},
\end{multline}
where $b(\mathbf{k},s)\equiv\int_{0}^{2\pi} a(\mathbf{k},\theta,s) \mathrm{d}\theta$. The RHS can be viewed as a differential operator $\mathcal{D}$ acting on $\tilde{f}_1$, where the kernel of the adjoint operator $\mathcal{D^\dagger}$ is  1. Thus, upon using the Fredholm Alternative theorem, setting the $X$-integral of the LHS to zero, one gets the conventional RT equation \eqref{ClassRT} with tumbling rate $\langle{\nu}\rangle$. Therefore, $D_{T\to0}=V^2/[2\nu_0 \exp{(\alpha^2/2)}]$, as previously anticipated.

The $T\to\infty$ limit  is also interesting and roughly corresponds to the experimentally fitted values for the \textit{E.~coli}, for which $\nu_0T\simeq12.3$. In this case a particle starts with a certain protein concentration (and hence a certain tumbling rate) as determined by $X$, which is then kept fixed at all times. 
The system is therefore equivalent to considering a ``polydisperse dilute fluid'', that is, a set of non-interacting particles, where each one has a fixed tumbling rate $\nu_i$ drawn from a continuous distribution. Thus the averaged diffusion coefficient is
\begin{equation}
D_{T\to\infty}\!=\!\Bigg\langle\frac{V^2}{2\nu_i}\Bigg\rangle\!=\!\frac{V^2}{2\nu_0}\!\int_{-\infty}^{\infty}\!\! e^{-(\alpha X+X^2/2)}\,\mathrm{d}X\!=\!\frac{V^2e^{\alpha^2/2}}{2 \nu_0 },
\label{D.smallalphav2}
\end{equation}
where we notice the opposite sign in the exponential argument in comparison to the $T\to0$ limit.

The two limits for $T$ and the small-$\alpha$ expansion \eqref{D.smallalpha} can be interpolated in a compact expression 
\begin{equation}
D_i=\frac{V^2 }{2 \nu_0}\exp \left[\frac{\alpha^2(\nu_0  T-1)}{2(\nu_0  T+1)}\right]. \label{DmodelX.interpolated}
\end{equation}
By changing $T$ between its two limits we change $\tau\equiv(\nu_0 T-1)/(\nu_0 T+1)$ in such a way that $\tau\in[-1,1]$ and, hence we have the bounds $D_{T\to0}\leq D\leq D_{T\to\infty}$. Simulations with different values of $\alpha$ and $T$ show that this interpolating expression is good for small $\alpha$ across distinct orders of $T$ (see Fig.~\ref{Fig4}b).

\section{Conclusions}
\label{conc}
Here we reviewed and extended general theoretical methods as well as performed simulations to investigate the approach to diffusion of run-and-tumble bacteria within four models: conventional run-and-tumble, partial reorientation, run-and-reverse with rotational diffusion, and stochastic tumbling rate. By focusing on the mean-squared displacement and on the excess kurtosis both analytically and computationally, we have extracted the effects of basic model parameters on how slowly diffusion is reached. The methods have been presented in a way that makes them easy to be translated into other models of particle dispersal. Although we have worked in 2D for the sake of simplicity, 3D generalizations should be straightforward to perform \cite{taktikos2013motility}. Furthermore, since many tracking experiments are performed in quasi-2D geometries~\cite{bechinger2016active}, our results are directly applicable.

For the conventional RT model with complete reorientation we obtained that the excess kurtosis approaches zero exponentially with a rate equal to the tumbling rate $\nu_0$. However, for the other models, new time scales appear, which can make the approach to the diffusive regime much slower. For the case of partial reorientation, the new time scales depend on the averages  $\sigma_1=\langle \cos\theta_{\mathrm{s}}\rangle$ and $\sigma_2=\langle \cos2\theta_{\mathrm{s}}\rangle$ of the  tumbling angle $\theta_{\mathrm{s}}$, and diverge when either of them approaches one. This happens when the $\theta_{\mathrm{s}}$ distribution is sharply peaked around both 0 and $180^{\circ}$. For the run-and-reverse model the new time scale is given by the inverse of the rotational diffusivity, $D_{\mathrm{r}}$. When $D_{\mathrm{r}}\ll\nu_0$, swimmers remain performing a one-dimensional random walk for a long time and transit slowly to the full diffusive motion. 
Finally, the stochastic tumbling rate model, which describes the dynamics of \textit{E.~coli}, is characterized by two parameters: the sensitivity $\alpha$ of the tumbling rate to the concentration fluctuations of a relevant protein and the memory time $T$ of this concentration fluctuations. Analytical results are obtained as an expansion for small $\alpha$, in which case long relaxation times, eventually diverging, are obtained for long memory times. Simulations are in excellent agreement. In this model we also compute the long-time diffusion coefficient, finding an expression valid for small $\alpha$ and any value of $T$.

Concomitantly, when the relaxation times grow, the same happens with the amplitude of the excess kurtosis, implying that the swimmer dispersion remains largely non-Gaussian for long times, even though the MSD can already increase linearly with time. The emergence of large relaxation times to reach the vanishing of the excess kurtosis implies that diffusion or reaction--diffusion equations cannot be used to describe bacterial dispersion at intermediate times and distances. Instead, kinetic theory or discrete element method simulations could be used. This becomes relevant in the design of microrobots for bioengineering applications \cite{ceylan2017mobile} which include, for example, killing pathogenous bacteria \cite{vilela2017microbots} or removing toxic heavy metals from contaminated water \cite{vilela2016graphene}.

By simulating with the experimentally obtained \textit{E.~coli} values for the partial reorientation model, $\nu_0 = \SI{1.0}{\second^{-1}}$ and $\langle\cos{\theta_{\mathrm{s}}}\rangle\approx0.33$ \cite{berg1993random}, we estimate that the time to reach an excess kurtosis $\gamma(t)$ such that $|\gamma(t)|=0.05$ is $t\approx\SI{71.2}{\second}$, a value that is independent of the swim speed $V$, as expected. A similar analysis can be done for the model with stochastic tumbling rate by using the previously mentioned values $T=\SI{19.0}{\second}$, $\nu_0 = \SI{0.65}{\second^{-1}}$, and 
$\alpha = 1.62$, and by setting $D_{\mathrm{r}}=0$ and $\sigma_1=0$~\cite{figueroa20183d}. This gives $|\gamma(t)|=0.05$ at $t\approx\SI{143.7}{\second}$.

For the intermediate time and length scales where the bacterial dispersion is not described by a diffusion equation, the computed expressions for the van Hove function [Eqs.~\eqref{rho.RTclasico}, \eqref{rhoks.aniso},  \eqref{rhoks.randr}, and \eqref{FinalRhoStocCase}] should be used as the Green function for the density evolution. Alternatively, moment equations derived from kinetic equations can be used (see~\cite{aranson2005pattern,saintillan2008instabilities,baskaran2008hydrodynamics} for examples of moment equations).

In future work we will use the methods employed here to compare how several types of interacting \cite{put2019non,soto2014run} swimmers approach diffusion. In particular, because of the richness imparted by polydispersity \cite{PabloPeter1,PabloPeter2,PabloPeter3,decastro2019}, fluid mixtures of interacting run-and-tumble particles with different swimming strategies will be studied. One might also want to tackle circularly propelled active particles and investigate similar associated phenomena including those dependent on the so-called reverse rotations of driven rigid bodies \cite{parisio2008reverse,de2014role}.

\section{Acknowledgments}
This research is supported by Fondecyt Grant No.\ 1180791 (R.S.) and by the Millennium Nucleus Physics of Active Mater of ANID (Chile).

\appendix*
\section{\uppercase{Simulations}}
\label{simu}

The four models considered in this paper have been simulated in 2D in the following direct way. First, units are chosen such that $V=\nu_0=1$. Each particle starts at the origin $\mathbf{r}=\mathbf{0}$ at $t=0$. In the time step $\Delta t=0.001$ the bacterium moves to a new position as determined by its fixed speed $V$ and varying orientation $\theta$, whose initial value is uniformly distributed between $0$ and $2\pi$. After each time step a new orientation is chosen if, and only if, a newly drawn random number between $0$ and $1$ is smaller than $\nu \Delta t$ where $\nu$ is the tumbling rate. The way the new $\theta$ is chosen as well as the values of $\nu$ depend on each model as follows.

For the first three models (Sec.\ \ref{cons}), a constant $\nu=\nu_0$ is taken. In the case of complete re-orientation, any new $\theta$ is drawn randomly between $0$ and $2\pi$, just like the initial orientation. In the partial re-orientation model, a new $\theta$ is determined from a tumbling angle $\theta_{\mathrm{s}}$ that is drawn from one of two distributions: in the first case the tumbling angle is uniformly distributed in the range $[-\Delta/2,\Delta/2]$, whereas in the second case the tumbling angle $\theta_{\mathrm{s}}$ is uniformly distributed in the ranges $[-\Delta/4,\Delta/4]$ and $[180^{\circ}-\Delta/4,180^{\circ}+\Delta/4]$. The third model is that of a run-and-reverse particle with rotational diffusion. In this case the ``reverse'' part of the model represents a change in the value of $\theta$ by an amount of $\pi$ at each tumble. Besides, as in any simple implementation of rotational diffusion, $\theta$ changes further at each time step (that is, not only at each tumble) by the amount $\sqrt{2 D_{r} \Delta t} \eta$ where $\eta$ is a normally distributed stochastic variable of mean $0$ and variance $1$. 

In the last model (Sec.\ \ref{stoc}) we remind that, although each new orientation is chosen between $0$ and $2\pi$ as in the complete reorientation model, the tumbling rate $\nu$ is no longer constant but rather it follows $\nu(X) = \nu_0 e^{\alpha X}$. The initial value of the stochastic variable $X$ is also normally distributed with mean $0$ and variance $1$. It then evolves at each time step following a simple Euler-like scheme appropriate to its stochastic differential equation (\ref{Xevol}), i.e., at each step $X$ changes by the amount $-X{\Delta t}/T+\sqrt{2{\Delta t}/T}\xi$, with $\xi$ normally distributed with mean $0$ and variance $1$, again. A large number of time steps is then performed for each bacterium until a chosen total physical time $t$ is reached. This whole time series is repeated $2\times10^5$ times in order to provide the averaged behavior of the bacteria. In the case of the log-log scale data presented in Fig.~\ref{Fig2}a the simulations were repeated $2\times10^6$, instead. For the numerical values of the diffusion coefficients, the total simulated time was $t=1000$, but virtually identical results can be obtained with $t=200$, which confirms that the MSD decays much more quickly than the excess kurtosis.

\bibliographystyle{unsrt}

\bibliography{Active.bib}

\end{document}